\documentclass[prd,aps,preprint,tightenlines]{revtex4}
\usepackage{epsfig}
\usepackage{amsmath}
\vspace{1in}
\usepackage{axodraw}
\preprint{DOE/ER/40762-258}
\preprint{UM-PP\#02-058}

\begin{document}
\title{Parton Distributions in Light-Cone Gauge: \\ 
Where Are the Final-State Interactions? }
\author{Xiangdong Ji}
\email{xji@physics.umd.edu}
\affiliation{Department of Physics,
University of Maryland,
College Park, Maryland 20742 }
\author{Feng Yuan}
\email{fyuan@physics.umd.edu}
\affiliation{Department of Physics,
University of Maryland,
College Park, Maryland 20742 }
\date{\today}          
\begin{abstract}

We show that the final-state interaction effects in the single target 
spin asymmetry discovered by Brodsky et al. can be reproduced 
by either a standard light-cone gauge definition of the 
parton distributions with a prescription of the 
light-cone singularities consistent with the 
light-cone gauge link, or a modified light-cone gauge
definition with a gauge link involving the gauge potential 
at the spatial infinity.

\vspace{10cm}
\end{abstract}
\maketitle
\newcommand{\be}{\begin{equation}}
\newcommand{\ee}{\end{equation}}
\newcommand{\ben}{\[}
\newcommand{\een}{\]}
\newcommand{\beqn}{\begin{eqnarray}}
\newcommand{\eeqn}{\end{eqnarray}}
\newcommand{\Tr}{{\rm Tr} }

Recently, Brodsky and collaborators have re-examined the significance
of the parton distributions measurable in deep-inelastic and other high-energy
scattering. They found that the final state interactions (FSI) between the
struck quark and target spectators yield distinct 
physical effects such as shadowing and single-spin asymmetry \cite{brodsky, brodsky1}.
These effects are of course contained in the light-cone gauge-link 
explicitly present in the definition of the parton distributions
in the non-singular gauges, in which the gauge potential vanishes
at the spacetime infinity. In the light-cone gauge, 
however, the gauge-link vanishes by choice, and 
the parton distributions in the conventional definition become
parton {\it densities} which are entirely determined 
by the ground state light-cone wave functions. Since, Brodsky et al. argue, 
the wave functions in principle do not encode any
information about the FSI, the usual parton distributions in the light-cone 
gauge must be deficient. When modified to take into account 
the FSI, they may no longer interpretable as the parton
densities. Hence, the provocative
claim \cite{brodsky}, ``structure functions are not parton probabilities."

This of course is not the first time that the light-cone gauge gives
us troubles. Over the years, it has been realized that although the
light-cone gauge simplifies many calculations significantly, it 
has associated light-cone singularities at $k^+=0$ which 
are hard to interpret
at times. In the earlier days of factorization proofs, it
was realized that these singularities prevent contour deformations necessary 
to cancel certain soft contributions \cite{bodwin}. The factorization
theorem for Drell-Yan process, for example, was proved
in covariant gauges \cite{sterman}. In the literature, a number 
of prescriptions have been suggested to regularize 
the singularities \cite{lightcone}. Physically light-cone
singularities arise from an incomplete 
gauge fixing for $k^+=0$ gluons, corresponding to 
degrees of freedom in choosing the boundary conditions for 
the gauge potentials at infinity. Physical results, of course, must be 
independent of how this additional gauge freedom is being fixed. 
Nevertheless, in the light-cone gauge, depending on ways residual
gauge fixing, certain 
interesting physics can and do migrate to $k^+=0$ or 
the spacetime infinity. Hence, one must be careful in 
differentiating interesting physical effects at $k^+=0$ such as
the FSI from light-cone artifacts. 

In this paper, we argue that the standard definition
of the parton distributions in the light-cone gauge requires
a unique prescription for the light-cone singularities---
the one that is constrained to reproduce the light-cone 
gauge link in the covariant gauge. In this case, the 
deep-inelastic scaling functions
are parton distributions, calculated in a wave function
containing the effects of final state interactions!
 Alternatively, if one demands 
the initial state wave function be real, then the usual light-cone 
gauge link in the definition of the parton 
distributions is incomplete. It ought to be 
supplemented with an additional contribution. 
In the non-singular gauges, 
this new eikonal factor does not contribute. But in the 
gauges such as the light-cone gauge where the gauge potential 
does not vanish asymptotically, the additional 
gauge link is responsible for the final
state interactions. We use the example of the single spin asymmetry 
discussed in Ref. \cite{brodsky1} to show that the FSI physics 
is faithfully reproduced in this approach.  

The transverse-momentum parton distribution is defined
in the literature as \cite{collins} 
\begin{eqnarray}
 f(x, k_\perp) &=& \frac{1}{2}
   \int {\frac{d\xi^-d^2\xi_\perp} {(2\pi)^3}} 
        e^{-i(\xi^-k^+-\vec{\xi}_\perp\cdot \vec{k}_\perp)} \nonumber \\
  && \times \langle P|\overline{\psi}(\xi^-,\xi_\perp)
     L^\dagger_{\xi_\perp}(\infty,\xi^-)\gamma^+L_0(\infty, 0)
\psi(0)| P\rangle \ , 
\label{density}
\end{eqnarray}
where a suitable regularization is needed for
gluon rapidity integrations \cite{cs}, and $L$ is the path-ordered 
light-cone gauge link 
\begin{equation}
  L_{\xi_\perp}(\infty,\xi^-) = P \exp\left(-ig\int^\infty_{\xi^-} 
        A^+(\xi^-,\xi_\perp) d\xi^-\right) \ . 
\end{equation}
In hard scattering, the gauge link $L(\infty,0)$ arises 
from the final state interactions between the struck
quark and the gluon field in the target spectators. Since the quark has
extremely high energy and is moving with the speed of light 
from, say, $z=0$ along the negative $z$ direction, 
the interaction can be represented by an eikonal line 
integrating over the whole history  
of the particle's trajectory, from $t=0$ to $t=\infty$, 
or $\xi^-=0$ to $\xi^-=(t-z)/\sqrt{2}=\infty$. In the literature where  
the factorization theorems are proved, the eikonal line
is not considered as final state interactions
because the loop integrations over the gluon momenta 
do not have pinched infrared singularities \cite{collins2}. As such,
one can deform the integrations to a contour on which  
the momentum exchanges are large, and so the eikonal phase  
is effectively accumulated almost instantly in the 
infinite momentum frame. In this paper, however, 
we follow Ref. \cite{brodsky,brodsky1} by refering 
the gauge link as (non-pinched) final-state interactions.

In the non-singular gauges, the above definition yields 
the correct gauge-invariant parton distributions. Indeed the 
final state interactions are properly represented by the light-cone 
gauge link \cite{collins2}. As an example, let us first calculate the 
asymmetrical part of the transverse momentum distribution in a nucleon 
due to its transverse polarization, the so-called Sivers function
\cite{sivers,others}. 
Since we are only interested in the matter of principle, 
we use the simple model introduced in Ref. \cite{brodsky1}
to study the polarization asymmetry discussed there.

Expanding Eq. (\ref{density}) to the first order in $g$ and dropping the
leading term which does not yield any asymmetry in the transverse momentum 
distribution, we have 
\begin{eqnarray}
  f^\perp_{1T}(x, k_\perp) &=& {\frac{1} {2}} \sum_n 
\int {\frac{d\xi^- d^2\xi_\perp } 
         {(2\pi)^3}} e^{-i(\xi^- k^+-\vec{\xi}_\perp \vec{k}_\perp)}
    \langle P|\overline{\psi}(\xi^-,\xi_\perp)|n\rangle \nonumber \\ 
    && \times \langle n|\left(-ie_1\int^\infty_0 A^+(\xi^-,0) d\xi^-
   \right) 
  \gamma^+\psi(0)|P\rangle
  + {\rm h. c.} \ , 
\end{eqnarray}
where $e_1$ is the charge of the struck quark and
$n$ represents the intermediate di-quark states. 
The notation $f_{\perp 1}$ follows Ref. \cite{boer} except 
the kinematic factor is included here. 
Introducing the interactions between the diquark and gluons, 
and the nucleon and quark-diquark, 
\begin{equation}
  {\cal  L}_{\rm int} = -g\overline{\psi}\psi_N \phi^*
         - ie_2 \phi^*(\stackrel{\rightarrow}{\partial^\mu}-
   \stackrel{\leftarrow}{\partial^\mu})\phi A_\mu + {\rm h. c.}
\end{equation}
where $\psi$ represents the quark field, $\psi_N$ the nucleon, 
$\phi$ the charged scalar diquark with charge $e_2$. 
At one-loop order, we have the following expression 
from Fig. 1, 
\begin{eqnarray}
   f^\perp_{1T}(x, k_\perp) &=& {\frac{-ig^2e_1e_2}
{4(2\pi)^3\Lambda(k_\perp^2)}}
       \int {\frac{d^4q} { (2\pi)^4}}
        {\overline U}(PS) (\not\!k + m) \gamma^+
        (\not\! k+\not\! q + m) U(PS) \nonumber \\
          && \times {\frac{2(1-x)-q^+ }{q^++i\epsilon}}
         {\frac{1}{ (k+q)^2-m^2+i\epsilon}}
         {\frac{1} {(P-k-q)^2-\lambda^2 + i\epsilon}}
         {\frac{1}{ q^2+i\epsilon}} + {\rm h. c.}, 
\end{eqnarray}
where $q^\mu$ is the gluon momentum. $M$, $m$ and $\lambda$ are the masses
of the nucleon, quark and diquark, respectively. 
$U(PS)$ is the on-shell spinor for the nucleon
with momentum $P$ and polarization $S$. $\Lambda(k_\perp^2)$ denotes
\begin{equation}
   \Lambda(k_\perp^2) = k_\perp^2 + x(1-x)\left(-M^2 + {\frac{m^2} {x}}
        + {\frac{\lambda^2}{ 1-x}}\right)\ .
\end{equation}
The $q^-$ integration can be done by the contour 
method.  Adding the hermitian conjugating contribution
results in taking the imaginary part of the eikonal propagator
$1/(q^++i\epsilon)$. We obtain, 
\begin{eqnarray}
   f^\perp_{1T}(x, k_\perp) &=& {\frac{-ig^2e_1e_2 }{ 8x(2\pi)^3\Lambda(k^2_\perp)}}
      (m+xM) \int {\frac{d^2 q_\perp }{(2\pi)^2}} {\rm Tr}\left[
    \gamma^\dagger (\not\! q_\perp-\not\! k_\perp)
        \not\! P \gamma_5 \not\! S\right] \nonumber\\
    &&\times  \frac{1}{ (q_\perp-k_\perp)^2}
    \frac{1}{M^2 - {\frac{q_\perp^2 + \lambda^2}{ (1-x)}}
   - {\frac{q_\perp^2 + m^2}{x}}} \ .
\end{eqnarray}
The integration over $\vec{q}_\perp$ yields,
\begin{equation}
    f^\perp_{1T} (x, k_\perp) = \frac{g^2 e_1e_2 }{(2\pi)^4}
     \frac{(1-x)(m+xM)}{4\Lambda(k^2_\perp)} \epsilon^{+\alpha\beta\gamma}
      k_{\perp\alpha} P_\beta S_\gamma
        \frac{1}{k_\perp^2}\ln \frac{\Lambda(k_\perp^2)}{ \Lambda(0)} \ , 
\label{dis}
\end{equation}        
where in the above formulas, we choose $P^+=1$.

\begin{figure}
\begin{center} \begin{picture}(230,90)(0,0)
\SetWidth{1.5}
\ArrowLine(0,0)(50,20)
\ArrowLine(180,20)(230,0)
\SetWidth{0.7}
\DashArrowLine(50,20)(100,20){2}
\DashArrowLine(100,20)(180,20){2}
\ArrowLine(50,20)(50,70)
\ArrowLine(50,71)(100,71)
\Line(50,69)(100,69)
\Gluon(100,71)(100,20){2.5}{6}
\LongArrow(100,50)(100,45)
\ArrowLine(180,70)(180,20)
\DashLine(140,90)(140,0){5}
\Vertex(100,70){2}
\Vertex(50,70){2}
\Vertex(100,20){2}
\Vertex(50,20){2}
\Vertex(180,20){2}
\Text(25,15)[b]{$P$}
\Text(45,45)[r]{$k+q$}
\Text(75,75)[b]{$q$}
\Text(205,15)[b]{$P$}
\Text(185,45)[l]{$k$}
\end{picture}  
\end{center}
\caption{One-loop contribution to the spin-dependent transverse momentum
distribution in the nucleon.}
\end{figure}
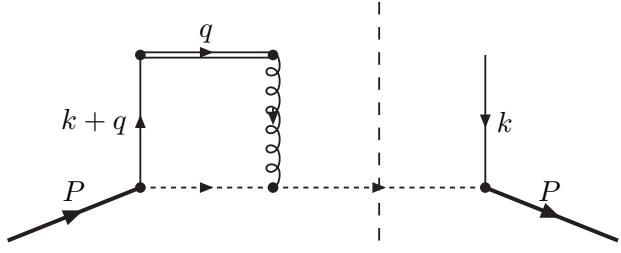

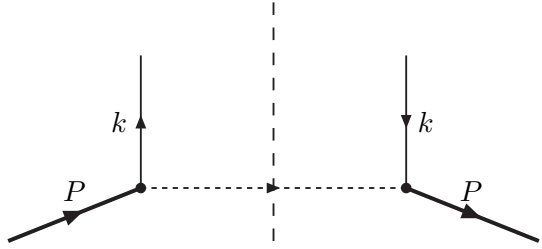
\begin{figure}
\begin{center} \begin{picture}(200,90)(0,0)
\SetWidth{1.5}
\ArrowLine(0,0)(50,20)
\ArrowLine(150,20)(200,0)
\SetWidth{0.7}
\DashArrowLine(50,20)(150,20){2}
\ArrowLine(50,20)(50,70)
\ArrowLine(150,70)(150,20)
\DashLine(100,90)(100,0){5}
\Text(25,15)[b]{$P$}
\Text(45,45)[r]{$k$}
\Text(175,15)[b]{$P$}
\Text(155,45)[l]{$k$}
\Vertex(50,20){2}
\Vertex(150,20){2}
\end{picture}  
\end{center}
\caption{Tree contribution to the spin-independent transverse momentum
distribution.}
\end{figure}

The spin-independent transverse-momentum distribution has a
contribution from the tree diagram shown in Fig. 2. A 
straightforward calculation leads to 
\begin{equation}
   f(x, k_\perp) = \frac{g^2 (1-x) \left[k_\perp^2+(xM+m)^2\right]}{2(2\pi)^3 
    \Lambda^2(k_\perp^2)}  \ . 
\end{equation}
The ratio of the spin-dependent and independent distributions is 
\begin{equation}
   \frac{f^\perp_{1T}}{f}(x, k_\perp) = \frac{e_1e_2(m+xM)}{4\pi P^+}\frac{\Lambda(k_\perp^2)}{ k_\perp^2
       + (xM+m)^2} \epsilon^{+\alpha\beta\gamma}
           k_{\perp\alpha} P_\beta S_\gamma \frac{1}{ k_\perp^2 }
      \ln  \frac{\Lambda(k_\perp^2)}{\Lambda(0)}\ . 
\end{equation}
This is the same as ${\cal P}_y$ in Ref. \cite{brodsky1} when 
$S^\mu=(0,0,0,1)$. 
Once again, the calculation demonstrates that the standard definition
of the parton distribution in the non-singular gauge
does take into account properly the effects of the final-state interactions
\cite{collins2}.  

In the light-cone gauge $A^+=0$, however, the gauge link 
$L$ vanishes. 
Where are the final state interactions? To find the answer, 
we consider all contributions to $f(x, k_\perp)$ at one-loop 
order in both Feynman and light-cone gauges. 
In the light-cone gauge, the gluon propagator has an extra term
\begin{equation}  
     \Delta D^{\mu\nu}(q) = {i\over q^2}
         {q^\mu n^\nu + q^\nu n^\mu \over q\cdot n} \ , 
\end{equation}
where ${1/q\cdot n}$ is singular at $q\cdot n=0$ and 
requires a regularization. In a scattering process, 
all the contributions coming from this extra term
cancel after using the Ward identity following
contracting the gluon momentum $q^\mu$ with the interaction
vertices, 
\begin{equation}
   {i\over \not\! k} \not\! q {i\over \not\! k + \not\! q}
  = i\left({i\over \not\! k} - {i\over \not\! k + \not\! q}\right) \ , 
\end{equation}
and the physical amplitude is independent of this
extra term (gauge-invariance) and hence the way this singularity
gets regularized. An explicit example of this can be found in Ref. 
\cite{brodsky}. 

In the case of parton distributions, however,
the cancellation is incomplete. Shown in Fig. 3a are the residual 
light-cone gauge contributions obtained by contracting 
$q^\mu$ from $\Delta D^{\mu\nu}$ with the vertices, where the 
shaded lines indicate absence of the propagator after applying 
the Ward identity. If gauge invariance holds, 
these contributions must exactly be equal to the corresponding
Feynman gauge contributions from the light-cone gauge link
shown in Fig. 3b. This happens only when the light-cone gauge
singularity is regularized according to the boundary condition 
imposed on the eikonal link. Therefore, we arrive at an interesting
conclusion that the parton distribution in Eq. (1) is valid
for the light-cone gauge only when the following 
light-cone gauge propagator is used   
\begin{equation}
     D^{\mu\nu}(q) = {-i\over q^2}
  \left(g^{\mu\nu} - {q^\mu n^\nu + q^\nu n^\mu \over q\cdot n+i\epsilon}
 \right) \ , 
\label{lc}
\end{equation}
where the direction of $q^\mu$ is toward the struck quark 
in its initial state. 
In other words, one now does not really has a freedom to choose
the regularization for the light-cone singularity, contrary
to the popular belief. The above propagator
can generate phases in the light-cone wave functions, 
reflecting the effects of the final state interactions, and thus 
the light-cone wave functions do not just contain the
structural information of the initial hadron. 

\begin{figure}
\SetWidth{0.7}
\begin{center} \begin{picture}(250,440)(0,0)
\SetOffset(0,330)
\SetWidth{1.5}
\ArrowLine(0,0)(20,20)
\ArrowLine(100,20)(120,0)
\SetWidth{0.7}
\DashArrowLine(20,20)(100,20){2}
\ArrowLine(20,20)(20,50)
\ArrowLine(20,50)(20,80)
\ArrowLine(100,80)(100,50)
\ArrowLine(100,50)(100,20)
\DashLine(60,90)(60,0){5}
\Gluon(20,50)(100,50){2.5}{10}
\Vertex(20,20){2}
\Vertex(100,20){2}
\Vertex(20,50){2}
\Vertex(100,50){2}
\Line(17,51)(23,53)
\Line(17,53)(23,55)
\Line(17,55)(23,57)
\Line(17,57)(23,59)
\Line(17,59)(23,61)
\Line(17,61)(23,63)
\Line(17,63)(23,65)
\Line(17,65)(23,67)
\Line(17,67)(23,69)
\Line(17,69)(23,71)
\Line(17,71)(23,73)
\Line(17,73)(23,75)
\Line(17,75)(23,77)
\Line(17,77)(23,79)

\SetOffset(-140,220)
\SetWidth{1.5}
\ArrowLine(140,0)(160,20)
\ArrowLine(240,20)(260,0)
\SetWidth{0.7}
\DashArrowLine(160,20)(240,20){2}
\ArrowLine(160,20)(160,50)
\ArrowLine(160,50)(160,80)
\ArrowLine(240,80)(240,20)
\DashLine(200,90)(200,0){5}
\Gluon(160,50)(220,20){2.5}{9}
\Vertex(160,20){2}
\Vertex(240,20){2}
\Vertex(160,50){2}
\Vertex(220,20){2}

\SetOffset(-280,110)
\SetWidth{1.5}
\ArrowLine(280,0)(300,20)
\ArrowLine(380,20)(400,0)
\SetWidth{0.7}
\DashArrowLine(300,20)(380,20){2}
\ArrowLine(300,20)(300,50)
\ArrowLine(300,50)(300,80)
\ArrowLine(380,80)(380,20)
\DashLine(340,90)(340,0){5}
\Gluon(300,50)(320,20){2.5}{4}
\Vertex(300,20){2}
\Vertex(380,20){2}
\Vertex(300,50){2}
\Vertex(320,20){2}

\SetOffset(-280,0)
\SetWidth{1.5}
\ArrowLine(280,0)(300,20)
\ArrowLine(380,20)(400,0)
\SetWidth{0.7}
\DashArrowLine(300,20)(380,20){2}
\ArrowLine(300,20)(300,35)
\ArrowLine(300,35)(300,65)
\ArrowLine(300,65)(300,80)
\ArrowLine(380,80)(380,20)
\DashLine(340,90)(340,0){5}
\GlueArc(300,50)(15,-90,90){2.5}{6}
\Vertex(300,20){2}
\Vertex(380,20){2}
\Vertex(300,35){2}
\Vertex(300,65){2}

\Line(297,65)(303,67)
\Line(297,67)(303,69)
\Line(297,69)(303,71)
\Line(297,71)(303,73)
\Line(297,73)(303,75)
\Line(297,75)(303,77)
\Line(297,77)(303,79)

\SetOffset(-280,110)
\Line(297,51)(303,53)
\Line(297,53)(303,55)
\Line(297,55)(303,57)
\Line(297,57)(303,59)
\Line(297,59)(303,61)
\Line(297,61)(303,63)
\Line(297,63)(303,65)
\Line(297,65)(303,67)
\Line(297,67)(303,69)
\Line(297,69)(303,71)
\Line(297,71)(303,73)
\Line(297,73)(303,75)
\Line(297,75)(303,77)
\Line(297,77)(303,79)

\SetOffset(-280,220)
\Line(297,51)(303,53)
\Line(297,53)(303,55)
\Line(297,55)(303,57)
\Line(297,57)(303,59)
\Line(297,59)(303,61)
\Line(297,61)(303,63)
\Line(297,63)(303,65)
\Line(297,65)(303,67)
\Line(297,67)(303,69)
\Line(297,69)(303,71)
\Line(297,71)(303,73)
\Line(297,73)(303,75)
\Line(297,75)(303,77)
\Line(297,77)(303,79)


\SetOffset(160,330)
\SetWidth{1.5}
\ArrowLine(0,0)(20,20)
\ArrowLine(100,20)(120,0)
\SetWidth{0.7}
\ArrowLine(20,81)(40,81)
\Line(20,79)(40,79)
\DashArrowLine(20,20)(100,20){2}
\ArrowLine(20,20)(20,80)
\ArrowLine(100,80)(100,50)
\ArrowLine(100,50)(100,20)
\DashLine(60,90)(60,0){5}
\Gluon(40,80)(100,50){2.5}{9}
\Vertex(20,20){2}
\Vertex(100,20){2}
\Vertex(40,80){2}
\Vertex(100,50){2}

\SetOffset(20,220)
\SetWidth{1.5}
\ArrowLine(140,0)(160,20)
\ArrowLine(240,20)(260,0)
\SetWidth{0.7}
\ArrowLine(160,81)(180,81)
\Line(160,79)(180,79)
\DashArrowLine(160,20)(240,20){2}
\ArrowLine(160,20)(160,80)
\ArrowLine(240,80)(240,20)
\DashLine(200,90)(200,0){5}
\Gluon(180,80)(220,20){2.5}{8}
\Vertex(160,20){2}
\Vertex(240,20){2}
\Vertex(180,80){2}
\Vertex(220,20){2}

\SetOffset(-120,110)
\SetWidth{1.5}
\ArrowLine(280,0)(300,20)
\ArrowLine(380,20)(400,0)
\SetWidth{0.7}
\ArrowLine(300,81)(320,81)
\Line(300,79)(320,79)
\DashArrowLine(300,20)(380,20){2}
\ArrowLine(300,20)(300,80)
\ArrowLine(380,80)(380,20)
\DashLine(340,90)(340,0){5}
\Gluon(320,80)(320,20){2.5}{7}
\Vertex(300,20){2}
\Vertex(380,20){2}
\Vertex(320,80){2}
\Vertex(320,20){2}

\SetOffset(-120,0)
\SetWidth{1.5}
\ArrowLine(280,0)(300,20)
\ArrowLine(380,20)(400,0)
\SetWidth{0.7}
\ArrowLine(300,81)(320,81)
\Line(300,79)(320,79)
\DashArrowLine(300,20)(380,20){2}
\ArrowLine(300,20)(300,50)
\ArrowLine(300,50)(300,80)
\ArrowLine(380,80)(380,20)
\DashLine(340,90)(340,0){5}
\Gluon(320,80)(300,50){2.5}{5}
\Vertex(300,20){2}
\Vertex(380,20){2}
\Vertex(320,80){2}
\Vertex(300,50){2}

\SetOffset(0,0)
\Text(60,-20)[]{(a)}
\Text(220,-20)[]{(b)}
\end{picture}  
\vskip 0.5cm
\end{center}
\caption{a). Extra contributions from the light-cone gauge 
parton distribution compared to the Feynman gauge distribution
without the light-cone link. b) The light-cone-link contributions
to the Feynman gauge distribution.}
\end{figure}
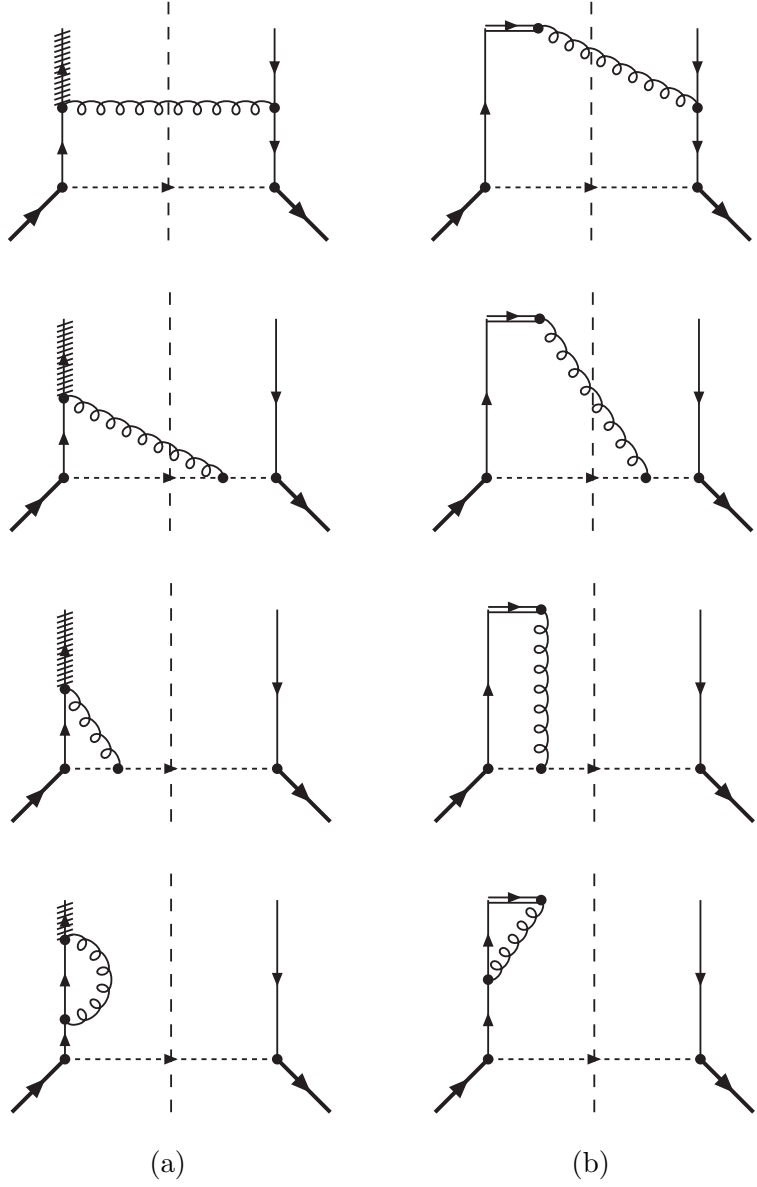

One can of course demand the light-cone wave functions
to be real and to contain only information about 
the initial hadron. This is achieved if
we treat $q^+$ as strictly real and $q^+=0$ is approached
symmetrically from both sides (principal-value prescription). 
In this case, we find that
the standard light-cone distribution has to be 
modified in the light-cone gauge, as the gauge potential
does not vanish at $\xi^-=\infty$. We show that this is 
the case for the example of single spin 
asymmetry discussed above.

Consider the Coulomb field generated by a charged particle moving 
at the speed of light. In the Feynman gauge, one has \cite{coulomb} 
\begin{equation}
              A^+ = -\frac{e}{4\pi}
       \delta(\xi^-) \ln \mu r_\perp, ~~A^-=0, ~~A_\perp = 0 \ . 
\end{equation}
The potential has support only at $\xi^-=0$. It has 
ultra-violet ($r_\perp\rightarrow 0$) and infrared divergences
($r_\perp\rightarrow \infty$). The former is related to 
the physical scale of a probe and the latter is due to the
vanishing photon mass. In the light-cone gauge, on the other
hand \cite{coulomb}, 
\begin{equation}
              A^+=A^-=0,~~ A_\perp = -\frac{e}{2\pi} \theta(\xi^-)\nabla \ln \mu r_\perp \ . 
\end{equation}
The perpendicular component $A_\perp$ does not vanish in the 
limit $\xi^-\rightarrow \infty$. This is the origin of the
spurious light-cone singularities. A regularization of the singularities 
corresponds to a choices of the boundary condition at $\xi^-=\pm \infty$. 
The anti-symmetric boundary condition for $A_\perp$ corresponds to 
a principle value regularization \cite{lightcone}.

If the gauge potential does not vanish at large $\xi^-$, 
it has a non-vanishing contribution to a gauge link at 
$\xi^-=\infty$. Therefore, the definition of the parton distribution 
in Eq. (\ref{density}) is no longer gauge invariant 
because the two light-cone links generated by $\psi(0)$ 
and $\psi(\xi^-, \xi_\perp)$ are not connected at $\xi^-=\infty$. If one makes a gauge 
transformation $U(\xi)$ which does not vanish at $\xi^-=\infty$, an
SU(3) matrix $U^\dagger(\xi^-=\infty, \xi_\perp)U(\xi^-=\infty,0)$ 
pops up in the distribution after the transformation. 
Therefore, Eq. (1) must be modified to a form that is
invariant under a singular gauge transformation. 

To motivate a modification, we consider again the motion of the struck
quark in deep-inelastic scattering. The quark has a large
$-$ momentum and thus it travels predominantly along the light-cone
$\xi^-$ direction. However, it also travels along the
transverse $\xi_\perp$ direction because 1) the quark inherits an intrinsic
transverse momentum from the nucleon; 2) it can exchange Coulomb
gluons with the target spectators. This transverse motion is justifiably
neglected in the eikonal approximation in the 
nonsingular gauges, and so the accumulation of the eikonal 
phase is entirely along the $\xi^-$ direction.

In the light-cone gauge the above approximation
is no longer valid. By definition, there is now no phase accumulation 
along the $\xi^-$ direction. However, since the phase is physical as
it determines the Coulomb scattering cross section, one is forced
to conclude that the phase must be accumulated through the
motion in the transverse direction. Since the particle is traveling 
mainly in the $\xi^-$ direction, the phase accumulation
in the transverse direction is {\it very slow}; appreciable phase accumulation 
happens at a very large $\xi^-$, or effectively $\xi^-=\infty$. 
Motivated by the above consideration, we modify the eikonal phase in 
Eq. (1) to,
\begin{equation}
     L_0(\infty,0)  \rightarrow  \Delta L=P\exp\left(-ig\int^\infty_0 
    d\xi_\perp\cdot A_\perp(\xi^-=\infty,
            \xi_\perp)\right)  \ , 
\end{equation}
where the path in the transverse direction is largely arbitrary. 
It is easy to see that the Coulomb field 
in the two gauges in Eqs. (11) and (12) yields the same phase factor
after integrating over the respective paths.

In the remainder of the paper, we will show that the rescattering
effect seen in the covariant gauge is easily reproduced in the
light-cone gauge. In fact, consider the
parton distribution in Eq. (1) 
with the gauge link $\Delta L$,  
  \begin{eqnarray}
  f^\perp_{1T}(x, k_\perp) &=& \frac{1}{2}\sum_n \int \frac{d\xi^- d^2\xi_\perp }{ 
         (2\pi)^3} e^{-i(\xi^- k^+-\vec{\xi}_\perp \vec{k}_\perp)}
    \langle P|\overline{\psi}(\xi^-,\xi_\perp)|n\rangle \nonumber \\ 
    && \times \langle n|\left(-ie_1\int^\infty_{\xi_\perp}d\xi_\perp' \cdot 
A_\perp (\infty,\xi_\perp')
   \right) \gamma^+
  \psi(0)|P\rangle
  + {\rm h. c.} \ . 
\end{eqnarray}
Going to the momentum space, we have
\begin{eqnarray}
   f^\perp_{1T}(x, k_\perp) &=& \frac{ig^2e_1e_2 }{4(2\pi)^3\Lambda(k_\perp)}
       \int \frac{d^4q }{(2\pi)^4}
        {\overline U}(PS) (\not\!k + m) \gamma^+
        (\not\! k+\not\! q + m) U(PS) (2(1-x)-q^+) \nonumber \\
          && \times \frac{e^{iq^+\infty}}{q^+}
         \frac{1}{ (k+q)^2-m^2+i\epsilon}
         \frac{1}{ (P-k-q)^2-\lambda^2 + i\epsilon}
         \frac{1}{ q^2+i\epsilon} + {\rm h. c.} 
\end{eqnarray}
where $1/q^+$ comes from the $n^-q^\perp/q^+$ term
in the light-cone propagator for the gluon. Using
\begin{equation}
      \lim_{L\rightarrow\infty} \frac{e^{iq^+L}}{ q^+} =
   i\pi\delta(q^+)\ , 
\label{limit}
\end{equation}
which is true in the sense of principal-valued distribution,
we recover the result in Eq. (\ref{dis}). 

A consistency check follows when replacing $q^+$ by $q^++i\epsilon$
in Eq. (\ref{limit}). The exponential factor
becomes $\exp(-\epsilon\infty)=0$. Therefore, if Eq. (\ref{lc}) is
used from the light-cone gauge propagator, the new gauge link
does not contribute. However, {\it the parton distributions defined
with the extra gauge link free one from choosing a
specific prescription for $1/q^+$}. Any prescriptions 
in fact will yield the same result. In particular, 
if one uses $q^+-i\epsilon$, both the hadron 
wave function and the extra link contain the effects of the 
final state interactions.

The gauge link $\Delta L$ can be derived in a similar way as the
ordinary light-cone gauge link \cite{furmanski}. It does not 
affect the Dokshitzer-Gribov-Lipatov-Alteralli-Parisi
evolution equations of the parton densities in the 
light-cone gauge because the evolution kernel is real. 
It is responsible for the shadowing effects (at amplitude level) 
considered in Ref. \cite{brodsky}. An account of these facts will 
be published elsewhere \cite{ji}.

To summarize, we have shown that the final state interactions 
can be taken into account in the light-cone gauge 
by either a gauge propagator chosen according to the 
physics of light-cone
gauge link in the usual parton distribution, or
an extra gauge link at $\xi^-=\infty$ in the parton 
distribution. In particular, we show that 
the single spin asymmetry discussed in Ref. \cite{brodsky1}
is recovered properly in the light-cone gauge. 
 
The authors thank S. Brodsky for a number of useful discussions. 
We also thank A. Belitsky, J. Collins, P. Hoyer, D. S. Hwang, 
S. Peigne, and J. W. Qiu for discussions and correspondences.
This work was supported by the U. S. Department of Energy via 
grants DE-FG02-93ER-40762.

\end{document}